\documentclass{PoS}

\usepackage{float}
\usepackage{amsmath}
\usepackage{amssymb}
\usepackage{amstext}
\usepackage{amsfonts}
\usepackage{graphicx}
\usepackage{multirow}
\usepackage{amssymb}
\usepackage{epsf}
\usepackage{epsfig}
\usepackage{epstopdf}
\usepackage[T1]{fontenc}
\usepackage{caption}
\usepackage{subcaption}

\newcommand{\ba}{\begin{eqnarray}}
\newcommand{\ea}{\end{eqnarray}}
\newcommand{\be}{\begin{equation}}
\newcommand{\ee}{\end{equation}}
\newcommand{\bd}{\begin{displaymath}}
\newcommand{\ed}{\end{displaymath}}
\newcommand{\een}{\nonumber\end{equation}}
\newcommand{\bea}{\begin{eqnarray}}
\newcommand{\eean}{\nonumber\end{eqnarray}}
\newcommand{\eea}{\end{eqnarray}}

\def\eq#1{Eq.~(\ref{#1})}

\def\fig#1{Fig.~\ref{#1}}

\newcommand{\gev}{\mathrm{GeV}}
\newcommand{\fm}{~\mathrm{fm}}

\newcommand{\sla}{\displaystyle{\not}}

\newcommand{\beq}{\begin{equation}}   
\newcommand{\eeq}{\end{equation}}   
\newcommand{\beqn}{\begin{eqnarray}}  
\newcommand{\eeqn}{\end{eqnarray}}

\def\mcO{{\mathcal O}}
\def\mcQ{{\mathcal Q}}

\def\psibar{\overline{\psi}}

			\def\mcE{{\mathcal E}}

			\def\mcL{{\mathcal L}}

			\def\mcO{{\mathcal O}}
			
			\def\mcQ{{\mathcal Q}}

			\def\ba{{\bf a}}

			\def\bd{{\bf d}}

\def\gev{\rm{GeV}}
\newcommand{\old}[1]{}

\title{Template Composite Dark Matter :\\ $SU(2)$ gauge theory with
  2 fundamental flavours.\\}

\ShortTitle{Template Composite Dark Matter}

\author{ \speaker{V. Drach}$^{a,b}$, A. Hietanen$\,^a$,
  C. Pica$\,^a$, J. Rantaharju$\,^a$,F. Sannino$\,^a$ \\
  \llap{$^a$}{$CP^3$-Origins \& the  Danish Institute for Advanced Study  DIAS, University of Southern Denmark, Campusvej 55, DK-5230 Odense M, Denmark\\}
  \llap{$^b$}{CERN, Physics Department, 1211 Geneva 23, Switzerland \\}
E-mail:\email{vincent.drach@cern.ch}
}

\abstract{We present a non perturbative study of SU(2) gauge theory with two
fundamental Dirac flavours. We discuss how the model can be used as a
template for composite Dark Matter (DM). We estimate one particular interaction of the DM candidate with the Standard Model : the interaction
through photon exchange computing the electric polarizability of the DM candidate.  Finally, we
briefly discuss the viability of the model given the present
experimental constraints.
\\
\\
{\textit{Preprint}} : CP3-Origins-2015-046 DNRF90 \& DIAS-2015-46}

\FullConference{The 33rd International Symposium on Lattice Field Theory\\
		14 -18 July 2015\\
		Kobe International Conference Center, Kobe, Japan*}

\begin{document}

\section{Introduction}

\begin{figure}[h!]
\centering
\includegraphics[width=0.25\textwidth]{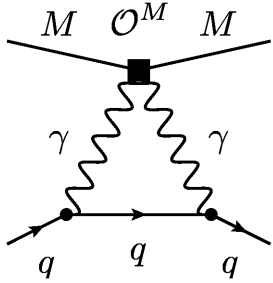}
\vspace*{-0.5cm}\captionof{figure}{Illustration of a two-photon interaction with quarks\cite{Frandsen:2012db}.}
\label{fig:diag}
\end{figure}

The  fact that  25\% of the energy content of the Universe must be
accounted for by Dark Matter (DM) is confirmed by
 several experiments. Many models have been
built to describe the Dark Matter sector of our Universe and are severely
constraints by experimental data obtained directly, indirectly, or
using colliders.  One interesting scenario is the one of Composite
Dark Matter where, as in baryonic sector, the most abundant particles
are ``composite''. In this work we consider one particular
realization of a composite dark matter model that breaks electroweak symmetry dynamically\cite{Cacciapaglia:2014uja}. 
The Dark matter candidate is electrically neutral and its interactions
with nucleon  receive contributions from  Higgs-exchange,   from
the electric dipole moment (which vanishes in the limit of degenerate
fermions), and at higher order in the operator expansion from a
two-photon exchange vertex, as illustrated in \fig{fig:diag}. While the two first
contributions have already been investigated in \cite{Hietanen:2013fya}, we
compute here the latter contribution and estimate the
cross section relevant for direct detection experiments. Note that
such an interaction has also be considered on the lattice in the context of
Stealth Dark Matter\cite{Appelquist:2015zfa}.

\vspace*{-0.8cm}

The model is based on $SU(2)$ gauge field theory with two fermions
in the fundamental representation. The Lagrangian reads in the continuum :
\be\label{eq:lagrangian}
\mcL=-\frac{1}{4} F^a_{\mu \nu} F^{a~\mu\nu} + \psibar \left ( i  \sla D - m\right) \psi,
\ee
where $\psi=(u,d)$ is a doublet of Dirac spinor fields transforming according to the
fundamental representation and can also be written as :

 \be
\mcL =-\frac{1}{4} F^a_{\mu \nu} F^{a~\mu\nu} + \psibar  i
  \sla D \psi + \frac{im}{2} \left[Q^T (-i \sigma_2) CE Q  +  \left(Q^T (-i \sigma_2) CE Q)\right)^\dagger\right]
\ee
where $\sigma_2$ acts on color indices and  $C$ is the
charge conjugation  matrix. Furthermore, we have defined :

\be
Q = \begin{pmatrix} u_L \\ d_L \\ -i\sigma_2 C \bar{u}_R^T \\
  -i\sigma_2
  C\bar{d}_R^T\end{pmatrix},\text{and}\quad E=\begin{pmatrix}0 &0
   & +1 & 0\\
0  &0  & 0 & +1\\
-1 &0  & 0 & 0\\
0  &-1 & 0 & 0\\
 \end{pmatrix}.
\ee
We have used $q_{L,R} = P_{L,R}
q,\bar{q}_{L,R} = \bar{q} P_{R,L}$ with  $P_L=\frac{1}{2}
(1-\gamma_5)$ and $P_R=\frac{1}{2} (1+\gamma_5)$.
The model  exhibits an $SU(4)$ flavour symmetry in the massless
limit. The 15 generators of the corresponding Lie
algebra will be denoted  $T^{a=1,\dots,15}$. 
After adding a mass term, the remnant flavour symmetry is the
group spanned by the algebra that preserves $E T^{a,T} + T^{a,T} E = 0$.  This
relation defines the  10-dimensional algebra of the $SP(4)$ group.
The chiral symmetry breaking pattern is thus expected to be $SU(4)
\rightarrow SP(4)$ leading to 5 Goldstone bosons.
The model has been investigated on the lattice in \cite{Hietanen:2014xca}, and  the
chiral symmetry breaking pattern  has been proven to be the expected
one \cite{Lewis:2011zb}.

As proposed in \cite{Cacciapaglia:2014uja}, the Lagrangian Eq.~(\ref{eq:lagrangian}) can be
embedded into the Standard Model in such a way that it interpolates
between composite Goldstone Higgs and Technicolor
models\cite{Ryttov:2008xe,Appelquist:1999dq}. The two limits are
parametrized by a single parameter whose value depends on contribution
from Standard Model loops.  In the so-called technicolor limit the
model breaks electroweak symmetry and $3$ of the Goldstone bosons
provide mass to W's and Z gauge bosons while the two remaining  Goldstone bosons
are stable and can be arranged in a electrically neutral complex scalar field,
denoted $\phi$,which is a Dark Matter candidate. As argued
in \cite{Cacciapaglia:2014uja}, the Dark Matter mass is generated via
loop diagrams involving electroweak bosons and top quarks, they predict that the mass is
proportional to the scale $f_\Pi = 246 \gev$, and we will thus
restrict ourselves to $m_{\phi} < 500 \gev$.
Note that the $SU(2)$ gauge theory with two fundamental fermions have
been used to build several Dark Matter models\cite{Appelquist:2015zfa,Detmold:2014qqa,Detmold:2014kba,Hochberg:2014kqa,Hansen:2015yaa}.

In this model the two quark $(U_L,D_L)$ are arranged in a $SU(2)_L$ doublet with hypercharge $0$, and the two remaining (Weyl) fermions are singlet of $SU(2)_L$ with hypercharge $\pm 1/2$. The electric charge matrix of the fermions is $\mcQ=\rm{diag}\left( \frac{1}{2},-\frac{1}{2},-\frac{1}{2},\frac{1}{2}\right)$.

\vspace*{-0.5cm}
\section{Electromagnetic properties}

Since the underlying fermions are not electrically neutral, the
effective theory describing the composite (Goldstone) Dark
Matter candidate is expected to generate a two-photon coupling. The goal of this work is to
investigate that particular contribution. The low energy coupling constant that enter into the process $\phi \phi \to \gamma \gamma$ is called the
polarizabity (measured in $\left[\fm \right]^3$) and enter in the
compton cross section which can be computed in
chiral perturbation theory. In order to perform a lattice
calculation, a different approach is followed.  As shown in
\cite{Tiburzi:2008ma}, the polarizability also
characterizes the response of the mass of a spin-0 neutral bound state
to a classical constant electromagnetic field according to the
following small field expansion : 
\be\label{eq:exp_small_field}
m^{(\mcE)} = m^{0}  + \frac{1}{2} 4\pi \alpha_E \mcE^2 + \dots\,,
\ee
where $\alpha_E$ is the electromagnetic polarizability.  The latter
 relation suggests a first principle approach to calculate the
 polarizability and has been used in QCD to determine polarizabilities
 of various hadrons, see for instance \cite{Detmold:2010ts}.
We briefly sketch the method in the following. In order to fulfill the 't Hooft
condition\cite{'tHooft:1979uj}, the electric field needs to be
quantized according to
\be\label{eq:E_field}
\mcE = (ea^2)^{-1} \frac{2\pi n}{Q N_t N_L}\equiv (ea^2)^{-1}  E\,,
\ee
where $e$ is related to the electromagnetic coupling constant
$\alpha=e^2/(4\pi)$, $a$ is the lattice spacing, $Q$ the charge of
fermion, $n$ is an integer, and $N_t,N_l$ are the temporal and spatial lattice
extent. Note furthermore that we have introduced the dimensionless
lattice field $E$.  The gauge links are then
multiplied by a position dependent field $U^{(E)}_\mu$ defined as follows :
\be
U_\mu^{(E)} = e^{i \mcQ A_\mu(x)} e^{iQ E N_t x_3 \delta_{\mu,4}
  \delta_{x_4 ,N_t-1}},\quad \text{where} \quad A_\mu(x)=(0,0,-Ex_4,0)\,,
\ee
$\mcQ$ is the electric charge of the fermion, $N_t$ is the lattice
temporal extent. In our calculation, the electromagnetic background field is included
only in the valence which leads to a systematic uncertainty. We
however expect the calculation to provide an
order of magnitude estimate of the polarizability.

In our model the dark matter candidate has the same quantum numbers as
the following diquark operator $u^T C \gamma_5 \sigma_2 d$. The mass
as function of the electric field can be estimated by computing the
effective mass corresponding to the following two-point function :
\be
C^{\mathcal{E}}_{\rm{2pt}}(t) = \sum_{\vec{x}}\langle \phi(x)
\phi^{\dagger}(0)\rangle =  C^{\mathcal{E},\pi^0}_{\rm{2pt},\rm{conn.}}(t)\,.
\ee 
The last equality -which relates $C^{\mathcal{E}}_{\rm{2pt}}(t)$ to
the connected part of the neutral pion two-point function- can be
derived using properties of the Wilson-Dirac operator for a two-color theory in a presence of a background field.

Note that by inspecting the corresponding effective theory, we
concluded that a  polarizability operator is generated at order
$\mcO(E^6)$ by an operator of the form $m^2_{\phi} F_{\mu\nu}F^{\mu\nu}
\phi^\ast \phi$. The polarizability is thus expected to vanish in the
chiral limit. Note  that  $F_{\mu\nu}F^{\mu\nu} \partial_\rho\phi   \partial^\rho\phi$ also appears in the
effective theory. The latter is not expected to vanish in the chiral
limit and  will thus dominate  the cross section for very light DM
mass. We will disregard that contribution in this preliminary
study.
\vspace*{-0.35cm}
\begin{figure}[h!]
\centering
\begin{minipage}{.5\textwidth}
  \centering
  \includegraphics[width=\linewidth]{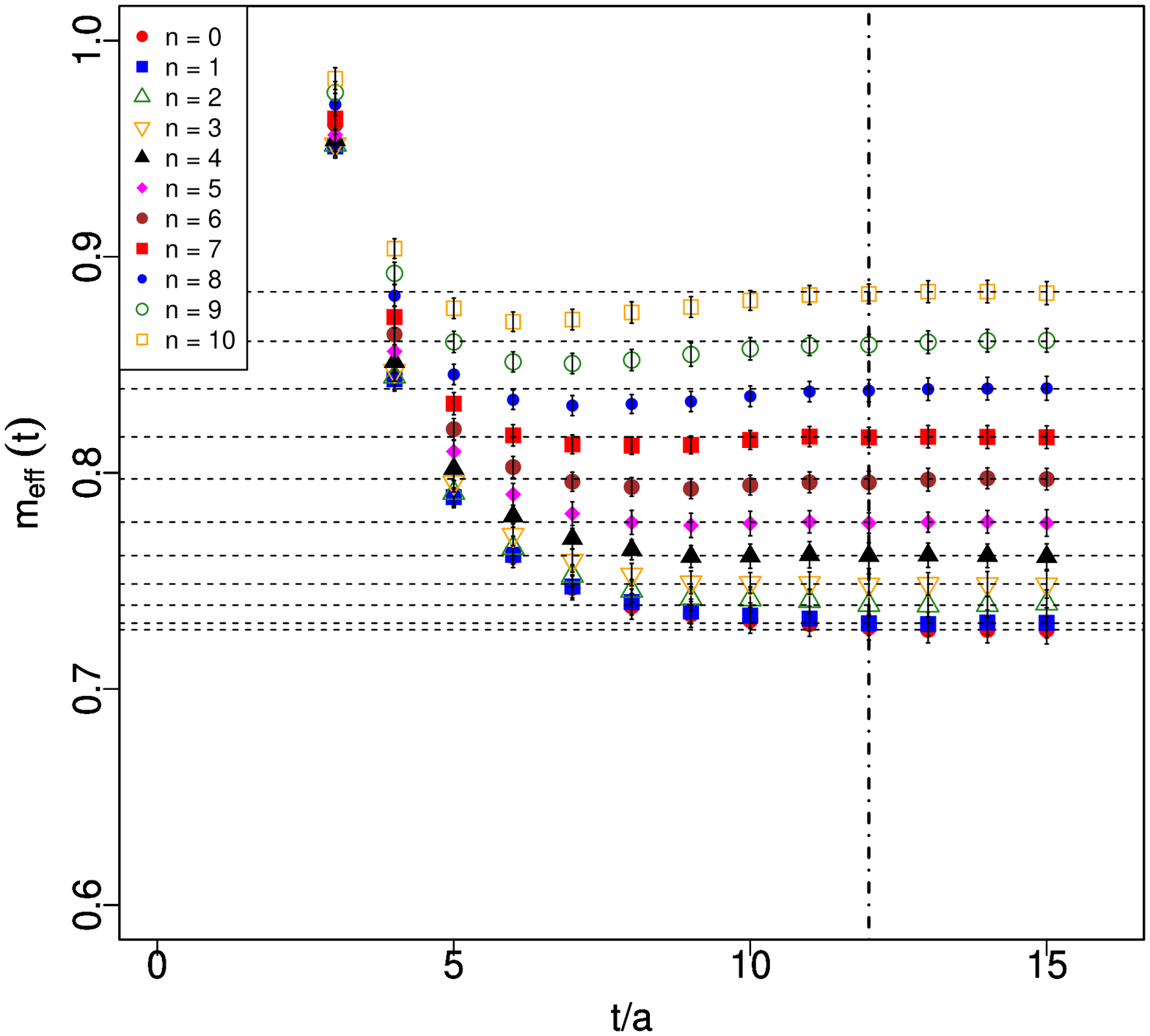}
  \captionof{figure}{Effective masses  for $\beta=2.2$, $L=16^3\times
    32$ and $m_0=-0.65$ for values of the electric field corresponding to
    $n=0,\dots,10$.}
  \label{fig:eff_mass}
\end{minipage}%
\hspace*{0.5cm}\begin{minipage}{.5\textwidth}
\centering
\vspace{0.35cm}
  \includegraphics[width=\linewidth]{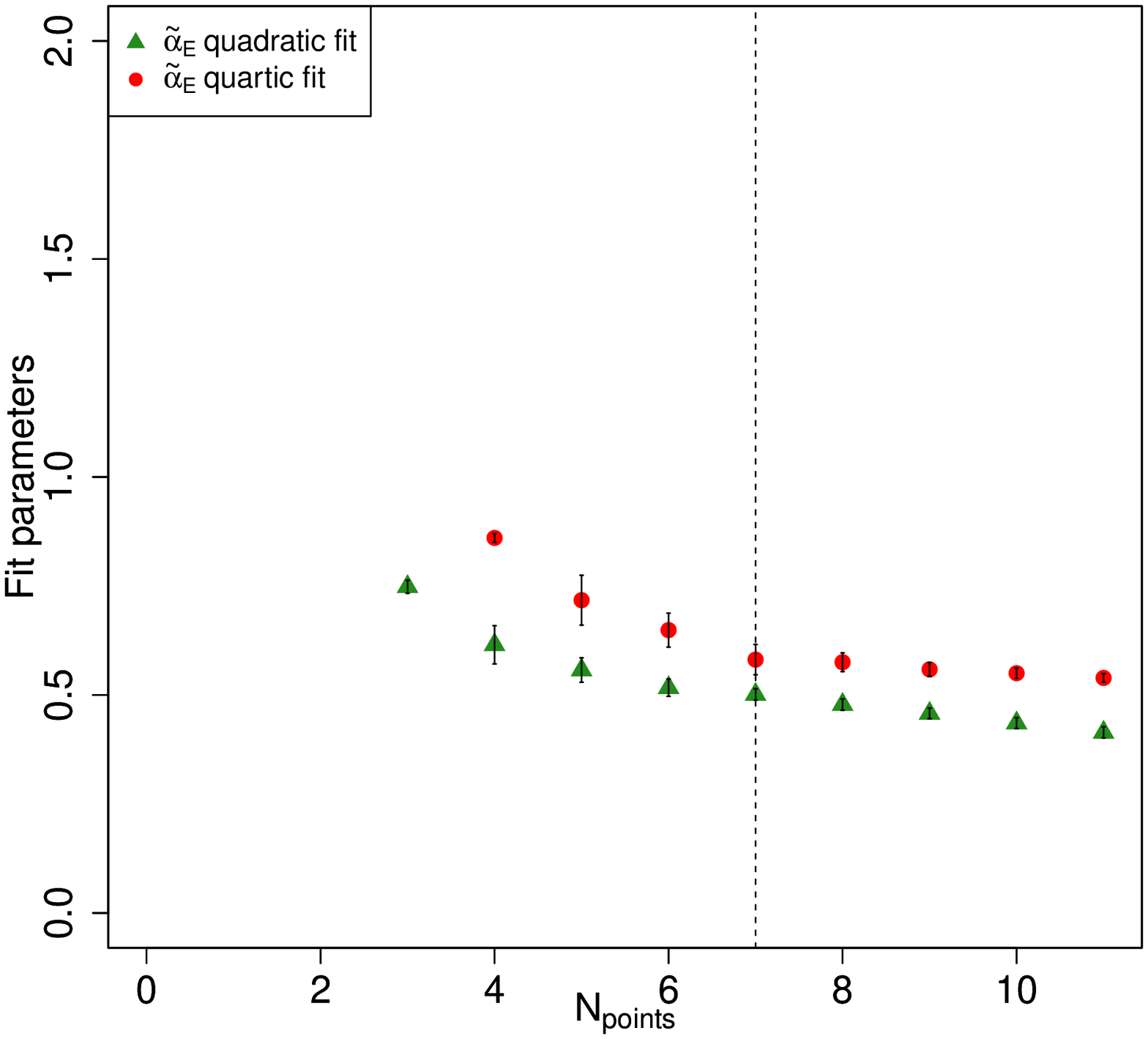}
  \captionof{figure}{$\widetilde{\alpha}_E$ assuming a quadratic or a quartic fit ans\"atz
as a function of the number of points included in the fit for $\beta=2.2$, $L=16^3\times
    32$ and $m_0=-0.65$ }
  \label{fig:pol_fit}
\end{minipage}
\end{figure}
\vspace*{-0.9cm}
\section{Results}
The simulation are performed using two flavour Wilson fermions with
the plaquette action. Two lattice spacing are used in this work and
the pseudoscalar decay constant $f_{\rm{PS}}$ is renormalized
perturbatively. Furthermore since the we are considering only the case
$\theta=\pi/2$ (technicolor limit), the scale is set by imposing
$F_\Pi=246 \gev$.
 
In \fig{fig:eff_mass} we show the effective mass defined by solving
the equation 
\be
\frac{C(t-a)}{C(t)} = \frac{e^{-m_{\rm{eff}}(t) (t-a)}
  +e^{-m_{\rm{eff}}(t) (T - (t-a))} }{e^{-m_{\rm{eff}}(t) t}
  +e^{-m_{\rm{eff}}(t) (T - t)}}\,,
\ee
for values of the electric field with $n=0,\dots,10$.  The results
show clear plateaus and the masses are obtained by fitting the effective
masses for $t/a$ larger than the vertical dotted line.  Once the mass
has been determined for each value of $n$, the dimensionless electric field
dependence is fitted according to :
\be
aM^{(E)} = aM^0 + \frac{1}{2} 4\pi \widetilde{\alpha}_E E^2 +
\mcO(E^4),\quad \rm{where}\quad \widetilde{\alpha}_E =
\frac{\alpha_E}{4\pi \alpha a^3}\,.
\ee
The relation between $\widetilde{\alpha}_E$ and $\alpha_E$ can be
derived by matching \eq{eq:exp_small_field} and using the definition
\eq{eq:E_field}. We show in \fig{fig:pol_fit} the fitted value of
$\widetilde{\alpha}_E$ assuming a quadratic or a quartic fit ans\"atz
as a function of the number of points included in the fit
($N_{\rm{points}})$. For $N_{\rm{points}}$, the two estimations of
$\widetilde{\alpha}_E$ are of the same order of
magnitude indicating that the contribution from the quartic coefficient is
small. Also, the value of $\widetilde{\alpha}_E$ does not
depend strongly on $N_{\rm{points}}$ and the fit is thus stable. Note furthermore that the expansion variable of
\eq{eq:exp_small_field} : $\frac{(e\mcE)^2}{m_{\rm{PS}}^4} \sim 0.03 n^2$ is small even for the smallest quark mass used in our setup.

\begin{figure}[h!]
\centering
\begin{minipage}{0.5\textwidth}
  \centering
  \includegraphics[width=\linewidth]{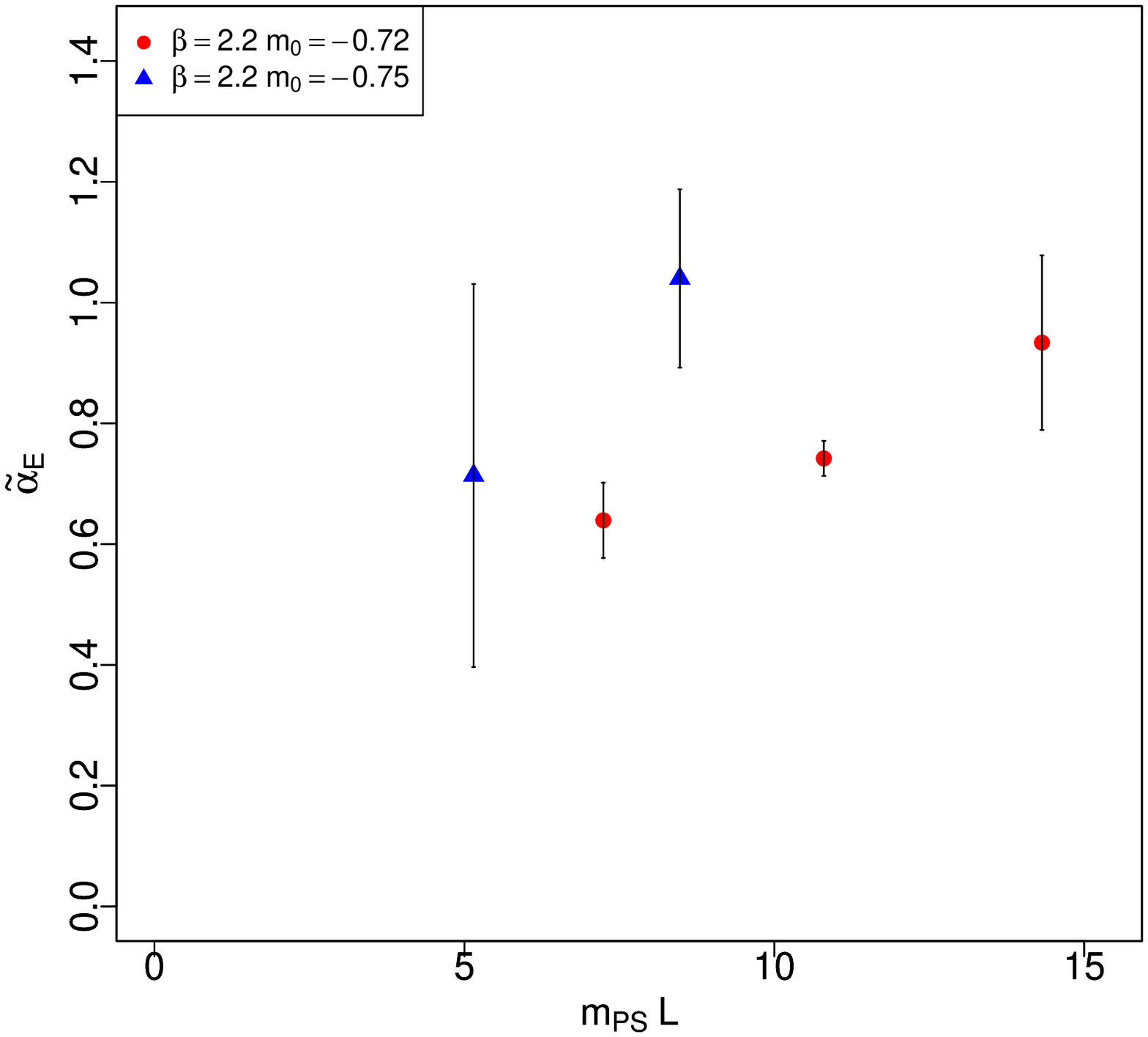}
  \captionof{figure}{$\widetilde{\alpha}_E$ as a function of $m_{\rm{PS}}L$ for two ensembles at
    $\beta=2.2$, including the lightest fermion mass used in our
    simulation.                                     }
  \label{fig:FV}
\end{minipage}%
\hspace*{0.5cm}\begin{minipage}{0.5\textwidth}
\centering
\vspace*{0.58cm}
  \includegraphics[width=\linewidth]{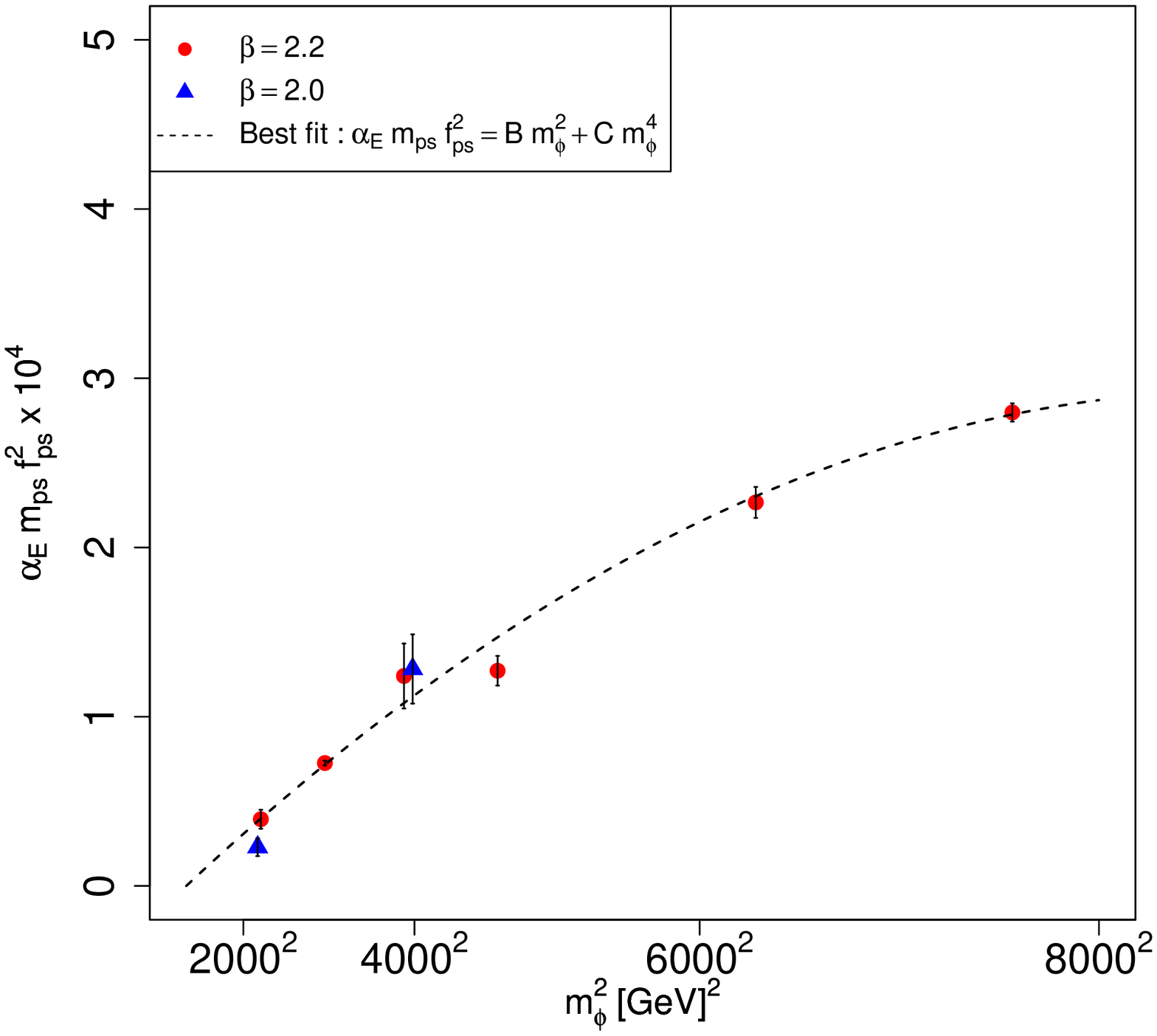}
  \captionof{figure}{$\alpha_E m_{\rm{PS}} f_{\rm{PS}}^2 \times 10^4$ as a
    function of $m_{\rm{PS}}^2 \equiv m_\phi^2$ at two different lattice
    spacing. A polynomial extrapolation including only $\beta=2.2$
    results is depicted by a dotted line.}
  \label{fig:xfit}
\end{minipage}
\end{figure}

In \fig{fig:FV}, we show the dependence of $\widetilde{\alpha}_E$ as a
function of $m_{\rm{PS}}L$ for two ensembles at fixed
$\beta=2.2$. Note that the lightest fermion mass is included in the
plot ($m_0=-0.75$). Since our results do not depend significantly of
the volume, we conclude that finite volume effects are
negligible in our results.
In \fig{fig:xfit}, we show the dimensionless quantity $\alpha_E m_{\rm{PS}} f_{\rm{PS}}^2$ as a function of  $m_{\rm{PS}}^2 \equiv m_\phi^2$ at two different lattice    spacing. We choose that particular combination because we expect it to cancel the leading order behaviour of $\alpha_E$. Note that the results obtained at $\beta=2.0$ show that
    our results are not significantly affected by lattice $\mcO(a)$
    effects. They are thus safely neglected in the following. We performed a polynomial fit of  $\alpha_E
m_{\rm{PS}} f_{\rm{PS}}^2$ imposing that $\alpha_E m_{\rm{PS}}
f_{\rm{PS}}^2$ vanishes in the chiral limit as required by the
effective field theory. The best fit value
obtained at fixed $\beta=2.2$ is depicted by a dotted line. Using the relation between $f_{\rm{PS}}$ and $m_{\rm{PS}}$ from
our previous work~\cite{Hietanen:2014xca}, we deduce a prediction for $\alpha_E(m_\phi^2)$.

Following \cite{Appelquist:2015zfa}, the effective interaction
Lagragian between DM and photons can be written $\mcL = \pi \alpha_E
F_{\mu\nu} F^{\mu\nu} \phi^\ast\phi$ and the cross section per nucleon
for a given target with atomic and mass number $(Z,A)$ can be written
as follows :
\be
\sigma_{\mathrm{nucleon}}(Z,A) = \frac{Z^4}{A^2} \frac{9\pi \alpha^2
  \mu_{n\phi}^2 (M_F^A)^2}{R^2} \alpha_E^2\,,
\ee
where $\mu_{n\phi}$ is the reduced mass, $\alpha$ is the
electromagnetic coupling constant, $R= 1.2 A^{1/3}$ and $ 1 < M_A^F <
3$ which enters in the nuclear part of the cross section. We refer to
\cite{Appelquist:2015zfa}, for a detailed discussion of the assumption
made to estimate the cross section.

\begin{figure}[h!]
\centering
\includegraphics[width=0.6\textwidth]{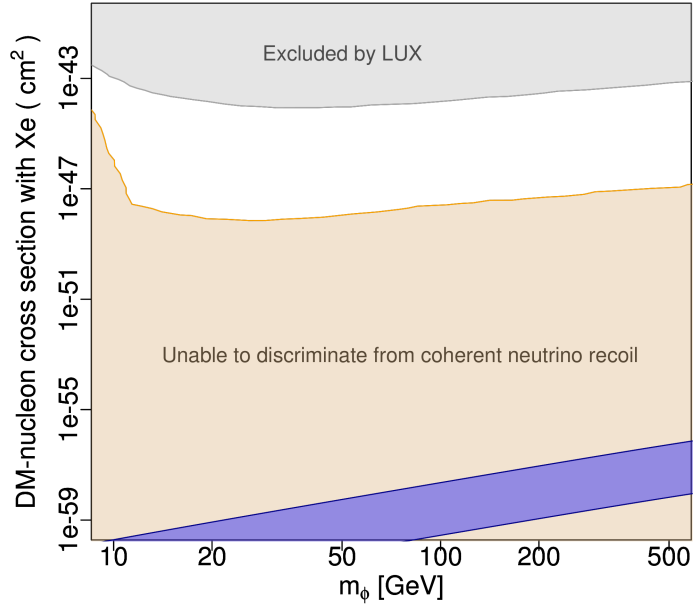}
\captionof{figure}{Prediction of the  nucleon-DM cross section per
  nucleon for  Xenon (blue band), constraints set by the LUX
  experiments\cite{Akerib:2013tjd}, and region were experiments are
  not able to discriminate from coherent neutrino recoil\cite{Billard:2013qya}.}
\label{fig:constraints}
\end{figure}

Using our prediction of $\alpha_E(m_{\phi}^2)$ we  plot by a blue band,
whose width is determined by the large uncertainty on the matrix
element $M_A^F$, the cross section per nucleon for Xenon in
\fig{fig:constraints}. In the figure, we also represented by a grey
area the latest constraints obtained by the LUX
experiments\cite{Akerib:2013tjd}. The orange filled area is the region of cross
section were experiments are not able to discriminate a nucleon-DM event
from a coherent neutrino recoil \cite{Billard:2013qya}. We conclude that, within our
assumption, the cross section due to two photon exchange is orders of magnitude too small to be detected.

\clearpage
\section{Conclusion}

We considered the interesting framework of a unified composite Higgs
and technicolor models. We argued that in the technicolor limit the
model features a DM candidate. Because of its composite nature, the DM
candidate is sensitive to electromagnetic interaction via the electric
polarizability of the DM. We performed a lattice calculation  in isolation of the SM  of the electric
polarizability using the background field method. We concluded that the
expected cross section is to small to be accessible via direct
detection. In future works we plan to investigate what happens beyond
the technicolor limit of the model and to study  the effect of  electroweak
corrections to the nucleon-DM cross section.

\vspace*{-0.5cm}
\section*{Acknowledgments}
\vspace*{-0.5cm}

This work was supported by the Danish National Research Foundation
DNRF:90 grant and by a Lundbeck Foundation Fellowship grant. The computing facilities were provided by the Danish Centre for Scientific Computing.

\end{document}